\documentclass[aps,pre,onecolumn,showpacs,showkeys,a4paper]{revtex4}
\input epsf  
\usepackage{graphicx} 
\usepackage{amsmath}
\usepackage{amssymb}
\usepackage{amscd}

\begin{document} 

\title{Noise-induced bifurcations, Multiscaling  and On-Off intermittency}
\author{S\'ebastien Auma\^ \i tre$^1$, Kirone Mallick$^2$, 
Fran\c cois P\'etr\'elis$^3$
}
 
\affiliation{$^1$ Service de Physique de l'Etat Condens\'e,  
Centre d'\'etude de Saclay, 91191 
Gif--sur--Yvette Cedex, France\\ 
$^2$Service de Physique Th\'eorique, Centre d'\'etude de Saclay, 91191 
Gif--sur--Yvette Cedex, France\\ 
$^3$Laboratoire de Physique Statistique, 
Ecole Normale Sup\'erieure, 24
rue Lhomond, 75005 Paris, France}

\begin{abstract}
  We present recent results on noise-induced transitions in a
 nonlinear oscillator with randomly modulated frequency.  The presence
 of stochastic perturbations  drastically  alters the  dynamical
 behaviour of the  oscillator~:  noise  can  wash out  a  global
 attractor but can also have a constructive role  by stabilizing   an
 unstable fixed point.  The random  oscillator   displays a rich
 phenomenology but remains  elementary enough to allow for exact
 calculations~:   this  system is thus  a  useful  paradigm for the
 study of  noise-induced  bifurcations and is an ideal testing ground
 for various mathematical techniques.  We  show that the phase  is
 determined by the sign of the  Lyapunov exponent (which  can be
 calculated non-perturbatively for white noise), and we  derive  the
 full  phase diagram  of  the system.  We  also  investigate the
 effect of time-correlations of the noise on the   phase diagram and
 show that  a   smooth random perturbation is  less efficient than
 white noise.  We  study the critical behaviour near the transition
 and explain why  noise-induced transitions often exhibit
 intermittency and multiscaling~:  these effects do not depend on
 the amplitude of the noise but rather  on  its  power spectrum. By
 increasing or  filtering out the low frequencies  of the noise,
 intermittency and multiscaling can be  enhanced or eliminated. 
 \end{abstract}

 \pacs{ 05.40.-a, 05.45.-a, 91.25.-r}
 \keywords{noise, stochastic  bifurcation, Lyapunov exponent,
  multiscaling, intermittency.}

\maketitle

\section{Introduction}

Most patterns  observed in nature are created by instabilities that
occur  in  an  uncontrolled noisy environment~:  convection in  the
atmospheric layers  and  in the mantle are subject  to  inhomogeneous
and fluctuating heat flux;  sand dunes are formed under winds  with
fluctuating directions and strengths. These fluctuations  usually
affect   the control  parameters driving the  instabilities, such as
the Rayleigh number which is proportional  to the imposed temperature
gradient in natural convection. These fluctuations  act
multiplicatively  on  the  unstable modes. In the same spirit, the
evolution of global quantities,  averaged under small turbulent
scales, can be represented  by a  nonlinear  equation with fluctuating
global transport coefficients   that  reflect  the 
complexity at small scales.  For instance, it has been shown that the temporal
evolution of the total heat flux in rotating convection  can be
described by  a non--linear equation with a multiplicative  noise
\cite{Neufeld}. The dynamo instability   that  describes the growth of
the magnetic field of the earth and  the stars because of   the motion
of conducting fluids  in their cores,   is usually  analyzed in
similar  terms~:~the magnetic field  is expected to  grow  at  large
scale,  forced by a  turbulent flow. Here again,  the parameters
controlling the growth rate of the  field are fluctuating \cite{Sweet}.

Since the theoretical predictions of Stratonovich \cite{Strato}, and
 the experimental works of Kawaboto, Kabashima and Tsuchiya
 \cite{Kawakubo}, it is well known that the
  phase diagram of a system  can be  drastically modified by the
 presence of noise  \cite{graham,lefever}.
 Because of stochastic fluctuations of the control
 parameter, the critical value of the threshold  may change and  noise
 can delay  or favor  a  phase transition 
\cite{vankampen,gardiner,risken,anishchenko}.
  When   a physical system   subject to noise   undergoes 
  bifurcations  into states  that have no deterministic
  counterparts,  the  stochastic phases  generated by randomness
  have    specific   characteristics
  (such as  their  scaling behavior and the associated
   critical exponents)  allowing  to   define 
  new universality classes \cite{munoz1}.

  A  straightforward  approach 
 to study  the effect of   noise
 on a bifurcation diagram    would consist in  analyzing 
  the  nonlinear  Langevin   equation
 that governs the system.  However,    the interplay
 between noise and nonlinearity  results  in subtle effects
 that make  nonlinear stochastic 
 differential equations hard to   handle \cite{vankampen,wax,arnold}.   
  One of the simplest   systems  that can be used as a paradigm for the
 study of noise-induced phase transitions  is the random frequency 
 oscillator (for a recent and detailed monograph devoted to
 this subject see
\cite{gitterman}). A deterministic oscillator with damping evolves 
 towards the equilibrium  state of minimal energy  that  represents
 the unique asymptotic state of the system. However, if  
 the frequency of the oscillator  is  a time-dependent variable, the
 behavior may change~: due to continuous energy injection into the
 system through the frequency variations,  the system may sustain
 non-zero oscillations even in the long time limit if the amplitude
 of the noise is large enough. 

 In this work,  we present  recent results 
  on noise-induced transitions
 in a nonlinear oscillator with randomly modulated frequency.
 The ouline of this article is as follows.

   In section~\ref{sec:LyapBif}, we show
 that an   explicit calculation of   the Lyapunov exponent   allows
 to determine exactly 
  the phase diagram of the  noisy Duffing  oscillator
   \cite{philkir1,philkir2}. 
  In the case of a  single-well  oscillator,  a large enough noise 
  destabilizes  the origin  (which
 is a fixed point for the  deterministic dynamics).  The double-well
 oscillator presents a richer structure~: the unstable fixed point
 becomes a  global attractor  of the stochastic 
 dynamics    when the amplitude of the noise  lies in 
 a  well-defined interval~: this is a  typical example of a reentrant
 transition (which was  observed  earlier in the more elaborate
 setting of spatially extended systems \cite{vdbtorral}).
 These properties are only qualitatively modified when
 the noise has temporal correlations.

 In section~\ref{sec:scaling}, we discuss the scaling exponents
 associated with the order parameter in the vicinity of the
 transition. For a  deterministic transition, it is well-known that
 the different  types of  bifurcations   can be classified by
 some characteristic exponents. We 
 show, in the stochastic case,
  that these exponents can be  modified by the noise~:
 the dynamical variable has a non-trivial scaling behaviour 
  and  its time series exhibits on-off intermittency.  Following 
 \cite{seb1,seb2},  we 
  explain that  multiscaling  and  intermittency
   are intimately linked with
 each other and that their existence is due to the presence
 of low frequency components in the noise spectrum. By using
 a random process with a spectral structure richer than
 that of white or Ornstein-Uhlenbeck noise, we show that
 the  deterministic exponents can be recovered by a suitable
 low-band flitering of the noise. This fact implies  that the
 concept of critical exponents  can be  ambiguous for noise-induced
  bifurcations  and is certainly not as  useful as it  is  for
  deterministic phase transitions.

 Section~\ref{sec:Poincare} presents  a discussion on perturbative
  perturbative expansion of the nonlinear Langevin equation. 
  Whereas most studies rely on 
 Fokker-Planck type
  evolution equations for the Probability Distribution 
 Function (PDF) in which the noise is  integrated out  by mapping  a
 stochastic  ordinary  differential  equation    into
 a deterministic partial differential  equation 
   in the phase space of the system, it is also possible to
 perform  a direct
 perturbative expansion of the nonlinear Langevin equation
 by using the classical  Poincar\'e-Lindstedt method \cite{luecke1,luecke2}.
 This  technique has the  advantage  as compared to the
 more traditional approaches of being applicable to 
   a noise with an arbitrary spectrum (whereas  Fokker-Planck equations
 are only valid for white-noise).   However, due to the presence
 of  low-frequencies  in the noise spectrum, 
 the perturbative  expansion breaks down and diverges.
 This divergence  is at the origin  of 
   the multiscaling behaviour and  of the anomalous scaling exponents. 
 The last section is devoted to concluding remarks.

\section{Stochastic Bifurcations of a random parametric oscillator}
\label{sec:LyapBif}

 In  this section, we   present  the phase diagram 
 of  a nonlinear oscillator
 with a randomly modulated  frequency. 
 The case when  the
 frequency is a periodic function of time 
  is a classical problem  known as the 
 Mathieu oscillator \cite{drazin,nayfeh,kevorkian}.
 The phase diagram of the (damped)  Mathieu oscillator   
 is  obtained by calculating the Floquet exponents (defined as  the
 characteristic growth rates of the amplitude of the system). This
  phase diagram   presents an alternance 
 of  stable regions in which the system 
 evolves towards its equilibrium  state and of unstable regions in which,
 because of parametric resonance, the amplitude of the oscillations
 grows  without bound. A nonlinear  term is needed 
  to saturate these oscillations.
  When the frequency of the pendulum is  a random process,
 the role of the Floquet exponents is taken over by the Lyapunov
 exponents. The system undergoes a bifurcation when the largest
 Lyapunov exponent,   defined as
  the  growth rate of the logarithm  of the  energy, changes its sign.
 Thus, the  Lyapunov  exponent vanishes on the 
  critical surface  that separates the phases in the parameter space.
  This criterion   involving  the sign of the  Lyapunov exponent
   resolves  the ambiguities that were found 
  in studies  of  the stability of higher  moments 
\cite{bourret,bourretFrisch,lindenberg3}
 and has  a firm mathematical basis  \cite{arnold}.
  In  recent works \cite{philkir1,philkir2}, we have 
  obtained  the exact phase diagram 
 of the random oscillator when the  frequency is a  Gaussian white noise. 
  In the last part of this section, we 
  discuss the effect of  non-vanishing time-correlation in the noise
  \cite{kirPeyneau}.

 \subsection{Instability induced   by noise: the single-well oscillator}

  The equation  for the amplitude $x$
 of a  random  Duffing oscillator   with a fluctuating
  frequency is 
\begin{equation}
 \ddot x  +  \gamma \dot x  + (\omega^2 +  \xi(t))\, x +  \lambda x^3 = 0 \, ,
 \label{randomOsc1}
\end{equation}
 where $ \gamma $ is the (positive)
 friction coefficient, $\omega$ the mean frequency
 and$\lambda$  the coefficient of the  cubic  non-linear term. The random 
 fluctuations of the frequency are 
 represented by a  Gaussian white noise  $\xi(t)$ of zero mean-value
 and of amplitude  ${\mathcal D}$ 
 \begin{eqnarray}
       \langle \xi(t)  \rangle &=&   0   \, ,\nonumber \\
 \langle \xi(t) \xi (t') \rangle &=& {\mathcal D} \, \delta(t - t')   \, .  
 \label{statxi}
\end{eqnarray} 
In this work, all stochastic differential equations are interpreted
according to Stratonovich calculus.  By  rewriting  time  and 
 amplitude in dimensionless units,
  $ t := \omega t$  and   $ x:= \lambda^{1/2} \omega^{-1} x,$
  Eq.~(\ref{randomOsc1})  becomes 
\begin{equation}
\frac{\mathrm{d}^2 x}{\mathrm{d} t^2} + 
\alpha  \, \frac{\mathrm{d} x}{\mathrm{d} t} + x + x^3 = x \, \Xi(t) ,
 \label{dissipDuff}
\end{equation}  
where  $\Xi(t)$ is a delta-correlated Gaussian  variable
 with vanishing mean value and with correlations given by 
  \begin{eqnarray}
 \langle \Xi(t) \Xi (t') \rangle &=& \Delta \, \delta(t - t')   \, .  
 \label{defXi}
\end{eqnarray} 
 The parameters 
\begin{equation}
  \alpha = \gamma/\omega \,\,\,\,\, \hbox{ and  }
      \,\,\,\,\,     \Delta = {\mathcal D}/\omega^3  \label{parametre}
\end{equation} 
correspond to dimensionless dissipation rate and to noise strength, 
respectively. 

 The equation~(\ref{randomOsc1}) has the origin of the phase space, 
  $ \dot x = x = 0$,  as  a {\it fixed point}. When the
 system is deterministic, {\it i.e.}, when   ${\mathcal D} = 0$,
 the  origin is a global
 attractor  for any  $\gamma > 0$~: for any initial condition
  we know that $x(t) \to 0$ and  $\dot x(t) \to 0$ when
 $t \to \infty$.

\begin{figure}
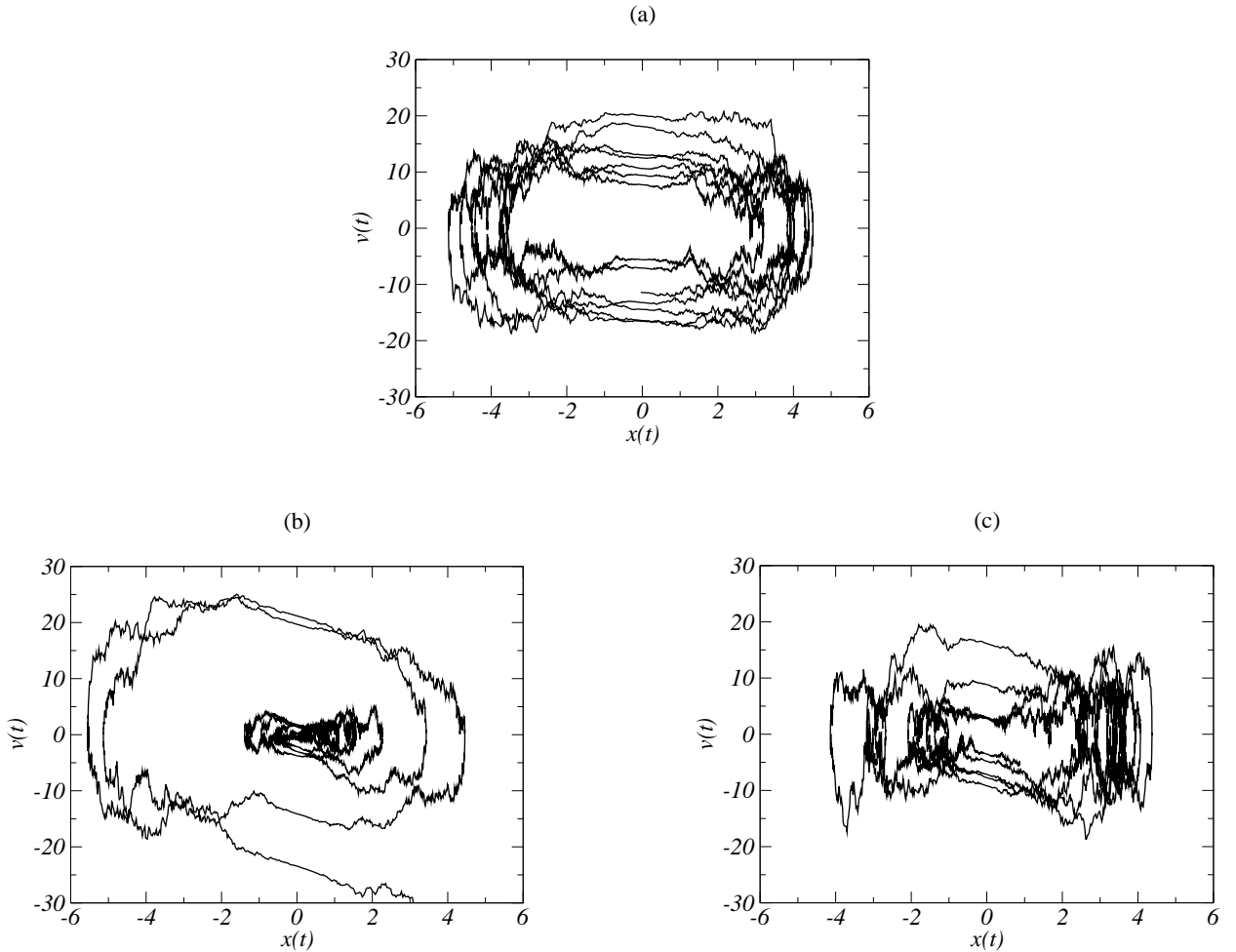

\includegraphics[height=6.0cm]{fig1a.eps}
\bigskip  
\bigskip

\includegraphics[height=6.0cm]{fig1b.eps}
{\hskip 2cm}
\includegraphics[height=6.0cm]{fig1c.eps}
\bigskip  

\caption{Phase plane plots of a typical trajectory of the noisy nonlinear
oscillator. Eq.~(\ref{dissipDuff}) is integrated numerically
with a time step $\delta t = 5 \ 10^{-4}$. (a) $\alpha = 0.5$,
$\Delta = 20$, $ t \le 100$; (b) $\alpha = 2.0 $,
$\Delta = 20$, $t \le 100$; (c) $\alpha = 2.0$,
$\Delta = 60$, $ t \le 100$. Early-time relaxation
towards noisy oscillations has been omitted for clarity in cases
(a) and (c). The trajectory spirals down towards the  origin 
in case (b).}
\label{fig:traj} 
\end{figure}

  However, in presence of noise, the asymptotic behaviour
 of the system becomes more complex.  
 In Fig.~\ref{fig:traj}, we present, for three values of the parameters,
a trajectory in the phase plane $(x, v = \dot x)$ characteristic
of the oscillator's behaviour, using the 
  numerical  one step collocation method advocated in 
\cite{mannella}. 
Initial conditions are chosen far from the origin, with an 
amplitude of order $1$. Noisy oscillations are observed
for small values of the damping parameter $\alpha$
[See Fig.~\ref{fig:traj}.a]. For  larger values  of $\alpha$  
the origin  becomes  a global attractor for the dynamics 
[See Fig.~\ref{fig:traj}.b].
Increasing the noise amplitude $\Delta$ at constant $\alpha$
makes the origin unstable again [See Fig.~\ref{fig:traj}.c].

 In other words, for the  stochastic oscillator,
 the following property is true~:  For a given $\gamma$ there is
 a  {\it  critical noise amplitude} $\mathcal{D}_c(\gamma)$ such that
\begin{eqnarray}
    &\mathcal{D} < \mathcal{D}_c(\gamma)& \,\,\, \,\,\,   \dot x, x \to 0
 \,\,\,  \hbox{  when }  t \to \infty   \nonumber \\
     &\mathcal{D} > \mathcal{D}_c(\gamma)&    \,\,\, \,\,\, 
 \dot x, x \to \hbox{ Oscillating Stationary State}    \nonumber
\end{eqnarray}
The main problem is  to calculate the value of  $\mathcal{D}_c$ as
 a function of  $\gamma$  and the other parameters of the problem.

\begin{figure}[th]
  \includegraphics[height=7.0cm]{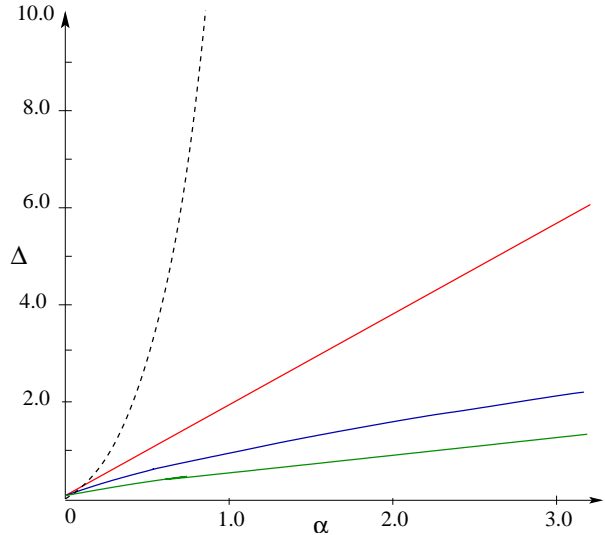}
 
\caption{Stability  diagram of  the moments
 of the single-well  oscillator. The full curves from top to bottom
 represent the stability lines of $\langle x^2 \rangle$,
  $\langle x^4 \rangle$ and $\langle x^6 \rangle$. The dashed curve
 is the line on which the Lyapunov exponent vanishes.} 
 \label{fig:moments}
\end{figure}

 For  random   dynamical systems,  it   was recognized early on that
  various `naive'  stability criteria \cite{bourret,bourretFrisch}, 
obtained by linearizing the  
dynamical equation  around the  origin, lead to ambiguous results.  
This  feature is in  contrast with  the  deterministic  case    for which
 the bifurcation threshold  is  obtained  without ambiguity
 by studying the eigenstates of  the linearized  equations
  \cite{manneville}. {\it  For the random oscillator,
  trying to determine the critical
 noise amplitude by studying the stability of the moments of the
 amplitude of the linearized equations leads to inconsistent results}.
   More precisely, consider the random harmonic oscillator,
 obtained by linearizing equation~(\ref{randomOsc1}) near the origin~:
\begin{equation}
 \ddot x  +  \gamma \dot x  + (\omega^2 +  \xi(t))\, x   = 0 \, .
 \label{randomOsc2}
\end{equation}
 The second moment $ \langle x^{2} \rangle$ of the amplitude
 of the linearized equation  converges to 0 in the long time
 limit if  
 $$  \frac{ \mathcal{D}}{\omega^3}  < 2  \frac{\gamma}{\omega} \, . $$
 The fourth moment  $ \langle x^{4} \rangle$ converges to 0  if
 $$  \frac{ \mathcal{D}}{\omega^3}  < 
   \frac{2\gamma}{3\omega} \,   \frac{3\gamma^2 + 4 \omega^2}
{3\gamma^2 + 2 \omega^2}\, . $$
 More generally, the moment  $ \langle x^{2p} \rangle$ is stable if
  $$  \frac{ \mathcal{D}}{\omega^3} 
 < \frac{2}{2p -1} \frac{\gamma}{\omega}
 {\mathcal F}_p \left( \frac{\gamma}{\omega} \right)  \, .  $$ 
 In fact,   it was conjectured in \cite{bourretFrisch}
 and proved in  \cite{lindenberg3} that,   in a linear oscillator 
 with arbitrarily small parametric noise, all moments beyond a certain 
 order grow exponentially  in the long time limit.  
  In figure~\ref{fig:moments}, we plot the stability diagram for the
 moments of order 2,4 and 6. We observe that the stability range
 becomes narrower when the  order of the moment becomes higher.
 This means that   any   criterion based on 
 finite mean  displacement,  momentum,  or energy of the  linearized dynamical
 equation   is not  adequate   to show the  stability
 of the initial problem and  the bifurcation 
threshold  of a  nonlinear random dynamical system cannot be determined  
simply   from  the moments of the linearized system.
In practice, the phase diagram  of the non-linear
 random oscillator  was  usually determined 
 in a perturbative manner by using weak noise expansions
 \cite{luecke1,luecke2,landa,drolet,landaMc}.

 The reason for 
 this failure is that in the linearized problem the statistics
 of the variable $x$ is dominated by  very large and very rare
 events that  induce the divergence of moments of high order.
 These rare events are suppressed by the nonlinearity and the
 bifurcation threshold becomes well defined. We 
  face the following  paradoxes~: (i)  the stability region of
 the origin  $\dot x = x =0$
  for  the non-linear equation has a well defined threshold for any value
 of the noise amplitude~:  below this threshold the origin is a global
 attractor of the dynamics. On the contrary the linearized equation
  does not have a clear-cut  bifurcation threshold. 
 (ii)  Close  to the origin the non-linear term
 is  irrelevant but  the  nonlinearity has to be taken into
 account to suppress the rare events that spoil the statistical
 behaviour of the system. 

 The problem is to find  the  correct 
  stability criterion for the non-linear equation that can  be formulated
 as in the deterministic case  on 
 the linear equation  and  that would allow   explicit  calculations.

  Such a  criterion does exist and is based on the Lyapunov 
 exponent that measures the growth rate of the random harmonic
 oscillator's energy 
\begin{equation}
\Lambda  = \lim_{t \to \infty} \frac{1}{2t} \langle \log E \rangle 
 \,\,\,\, \hbox{ with } \,\,\, E = \frac{\dot x^2}{2} + \omega^2
  \frac{x^2}{2} \, , 
\label{eq:defLyap}
\end{equation}
 where $x(t)$ is a solution of  equation~(\ref{randomOsc2}).
  It has been  shown (see \cite{philkir1}  and references therein)
 that when the Lyapunov  
 exponent is negative the  Fokker-Planck equation has a unique
 stationary solution which is the Dirac delta function at the origin
 of the phase space. This means that the  origin is a stable global
 absorbing state.   However, when the Lyapunov  
 exponent is positive and extended,   stationary probability distribution
 function exists and  describes an oscillatory asymptotic state
 of the nonlinear random oscillator. These features are reminiscent
  of Anderson localization and there exists indeed a mapping between
 the random oscillator and a 1d  localization  model \cite{tessieri,jmluck}.

  The equation of the transition line that determines
 the stability of the origin for the nonlinear
 oscillator with a random phase  is therefore given by
 \begin{equation}
   \boxed{  \,\,\,\,   \,\,\,\, 
  \Lambda(\gamma, \omega, \mathcal{D})  = 0    \, .  \,\,\,\, \,\,\,\,  } 
 \label{eq:stabline}
 \end{equation}
  The following  exact   closed  formula for the
 Lyapunov exponent  of the system 
 can be derived  when the random frequency modulation  is a
 Gaussian  white noise
 \cite{hansel,imkeller,philkir1}~:
\begin{eqnarray}
 \Lambda  &=&  \frac{\gamma}{2} \left( \frac{ \int_0^\infty \sqrt{u} \,\, 
  \rm{e}^{-\psi(u)}  \rm{d}u }{ \int_0^\infty \frac{\rm{d}u}{  \sqrt{u}}
 \,\,  \rm{e}^{-\psi(u)}  }  -1    \right) \label{eq:Lyap} \,  \\
 \hbox{  with }
  \psi(u)   &=& \frac{2\gamma^3}{ \mathcal{D} } 
  \left\{  \left(\frac{\omega^2}{\gamma^2} - \frac{1}{4}   \right) u  
    + \frac{u^3}{12}  \right\}   \, . 
\end{eqnarray}

\bigskip

\bigskip

 \begin{figure}[th]
  \includegraphics[height=6.0cm]{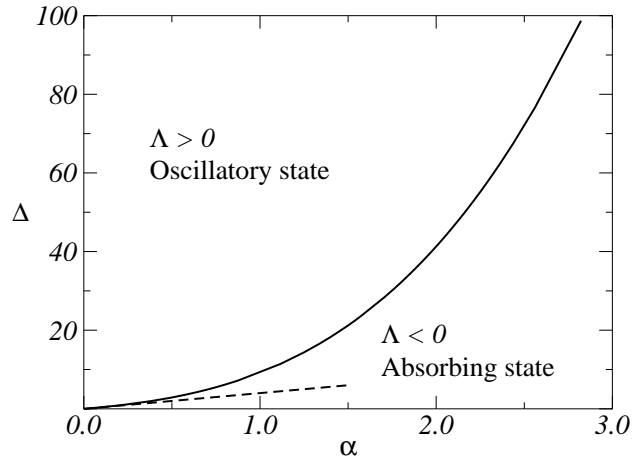}
 \bigskip
\bigskip
\caption{Phase diagram of the single-well  oscillator
  with multiplicative noise.
The critical curve separates an absorbing and an oscillatory phase
in the $(\alpha,\Delta)$ plane. The dashed line represents the 
small noise approximation (see text).} 
 \label{fig:diagphase1}
\end{figure} 

   Inserting this formula in the stability
 criterion~(\ref{eq:stabline})  and using  dimensionless variables
  we obtain  the equation of the transition line in the ($\alpha$,  $\Delta$)
 plane.   For a given value
 of $\Delta$, the  critical value of the damping 
$\alpha = \alpha_c(\Delta)$  for which the Lyapunov exponent vanishes
  is given by 
\begin{equation}
\alpha_c   =  {\displaystyle { \frac {\int_0^{+\infty} 
\mathrm{d}u \; {\sqrt u} \; 
  \exp\left\{-\frac{2}{\Delta} \left(  ( 1 - \frac{\alpha_c^2}{4} ) u
   + \frac{u^3}{12} \right) \right\}  }
 {\int_0^{+\infty} \frac{\mathrm{d}u}{\sqrt u}
  \exp\left\{-\frac{2}{\Delta} \left(  ( 1 - \frac{\alpha_c^2}{4} ) u
   + \frac{u^3}{12} \right) \right\}  } }. }
\label{alphac}
\end{equation} 
The critical curve 
$\alpha = \alpha_c(\Delta)$ 
is represented in Fig.~\ref{fig:diagphase1}.
It separates two regions in parameter space: for $ \alpha <  \alpha_c$
 (resp. $\alpha >  \alpha_c$)
 the Lyapunov exponent is positive (resp. negative),
the stationary PDF is an extended function (resp. a delta
distribution) of the energy and  the origin is unstable (resp. stable).
 For small values of the noise amplitude, it can be shown that
 the exact formula~(\ref{alphac}) reduces at first order to
 the linear relation $\alpha_c  = \frac{\Delta  }{4},$ 
  in agreement with previous perturbative calculations. 
 For large values of the noise,  we obtain 
 $ \alpha_c \simeq  0.656 \Delta^{1/3}  \, .$ We emphasize
 that the phase diagram drawn  in Fig.~\ref{fig:diagphase1} is exact
 for all values of the noise amplitude $\Delta$ and the damping
 parameter $\alpha$.

 \subsection{Stabilisation by noise: the double-well oscillator} 

 In a  classical calculation,  Kapitza  (1951)   showed  that 
the unstable upright position of an inverted pendulum is  stabilised
 if  its suspension axis undergoes sinusoidal vibrations
of high enough frequency.  Analytical
derivations of the stability limit are based on
perturbative approaches, {\it i.e.}, in 
 the limit of  small forcing  amplitudes \cite{landau,nayfeh,barma}.

 When the sinusoidal vibrations of the suspension axis are replaced
 by a white noise, an {\it exact}, non-perturbative, stability
  analysis can be performed  \cite{philkir2}. In particular, for 
 the stochastic  inverted pendulum,
  we have shown  
 that the unstable fixed point can be stabilized by noise and  have discovered
 the existence of a noise-induced reentrant transition.

 The general equation for the Duffing oscillator subject to multiplicative
 noise is given by~:
\begin{equation}
\ddot x  +  \gamma  \dot x     - 
 (\mu +  \xi(t))\, x + x^3 = 0  \, .
 \label{eq:doublewell}
 \end{equation}
 When $\mu < 0$,  the non-linear potential has a single-well.
 This case is the same as the one discussed in the previous section~:
 the origin is  deterministically a global attractor
 that  can become unstable  in presence of noise. When  $\mu >  0$,  
  the oscillator is subject to a double-well potential and 
 the   origin is deterministically unstable. 

 \begin{figure}[th]
  \includegraphics[height=5.0cm]{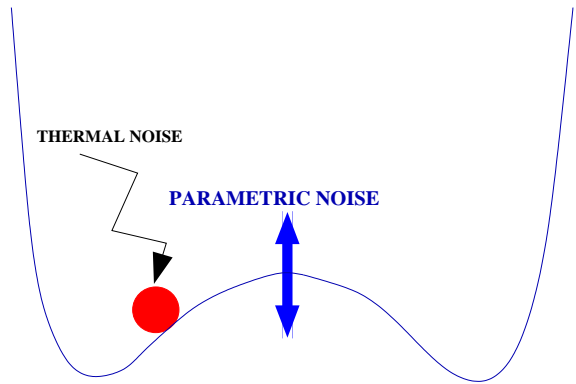}
   \caption{The double-well oscillator with parametric noise}
  \label{fig:2well}
\end{figure} 

 In presence of noise, 
   the correct criterion for   stability analysis
  is again  based on the sign of the Lyapunov 
 exponent.  In order to  take into account both  possible signs 
 of $\mu$, it is  useful to define  the following  set
 of   dimensionless parameters~:
\begin{equation}
 \alpha =  \frac{\mu}{\gamma^2}
    \,\, \,\, \hbox{ and }  \Delta   =  
 \frac{ {\mathcal D} }{ \gamma^3 }  \, . 
  \label{defalpha}
\end{equation}

\begin{figure}[th]
  \includegraphics[height=6.5cm]{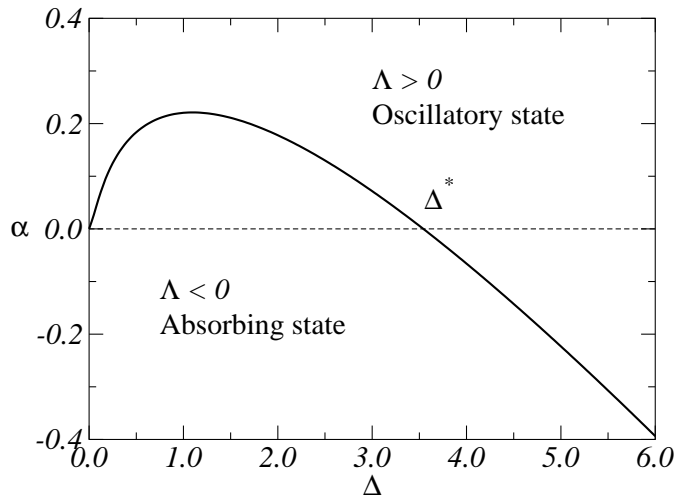}
 \bigskip
 \bigskip
  \caption{Phase diagram of the double-well oscillator subject
 to multiplicative Gaussian white noise.
 The solid line is the locus  
where $\Lambda(\alpha,\Delta) = 0$. The bifurcation line $\alpha = 0$ 
of the noiseless dynamical system is drawn for comparison (dotted line).
For $\Delta \le \Delta^* \simeq 3.55$ (resp. $\Delta \ge \Delta^*$), 
the origin is stabilised (resp. destabilised) by the stochastic
forcing in the range $0 \le \alpha \le \alpha_c(\Delta)$
(resp. $\alpha_c(\Delta) \le \alpha \le 0$).}
  \label{fig:diagphase2}
\end{figure}

  An exact calculation of the Lyapunov exponent \cite{philkir2}
  then allows to draw the phase diagram represented in 
 figure~(\ref{fig:diagphase2}).

 We observe 
 that when  $\alpha < 0.21\ldots$,  there exits a range
 of noise amplitudes for which the origin is {\it stabilized
 by noise}~: for  $\Delta$ such that 
  $\Delta_1(\alpha) < \Delta < \Delta_2(\alpha)$ 
  the origin becomes  an attractive fixed point of the stochastic dynamics.
  The values $\Delta_1(\alpha)$ and   $\Delta_2(\alpha)$
  that determine the stability interval are known analyticaly. We emphasize
 that these functions can not be calculated perturbatively by
 using a small noise expansion when  $\alpha$ is finite. For 
 $\alpha > 0.21\ldots$,  the stability interval does not exist anymore~:
 the origin is always unstable and
 the non-equilibrium stationary state exhibits an oscillatory
 behaviour.

 \begin{figure}[th]
  \includegraphics[height=2.2cm]{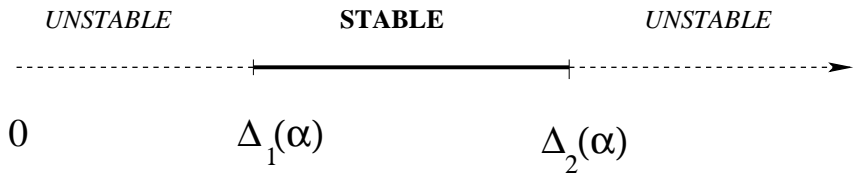}

   \caption{Stability range  for  the double-well oscillator~:
 when $\alpha < 0.21$, the unstable equilibrium point is
 stabilized by noise for noise amplitudes between $\Delta_1$
 and $\Delta_2$.}
  \label{fig:Stab}
\end{figure} 

 \subsection{The effect of colored noise}

  We now discuss the phase diagram of an 
 oscillator whose frequency is a random process with finite
 time memory. More precisely, we  consider  the case of 
 an  Ornstein-Uhlenbeck noise $x(t)$ of   correlation time $\tau$~:
 \begin{eqnarray}
 \langle \xi(t) \xi(t') \rangle &=&  \frac{{\mathcal D}_1}{2 \tau}
   \exp\left( - |t - t'|/\tau    \right) \,   \, .  
 \label{statxiOU}
\end{eqnarray} 
  When $\tau \to 0$, the process 
 $\xi(t)$ becomes identical to the white noise. The Lyapunov
 exponent of the system $\Lambda(\omega, \gamma, {\mathcal D}, \tau)$
  becomes now a function of $\tau$ also.

\begin{figure}[th]
 \centerline{\includegraphics*[width=0.375\textwidth, angle =-90]
 {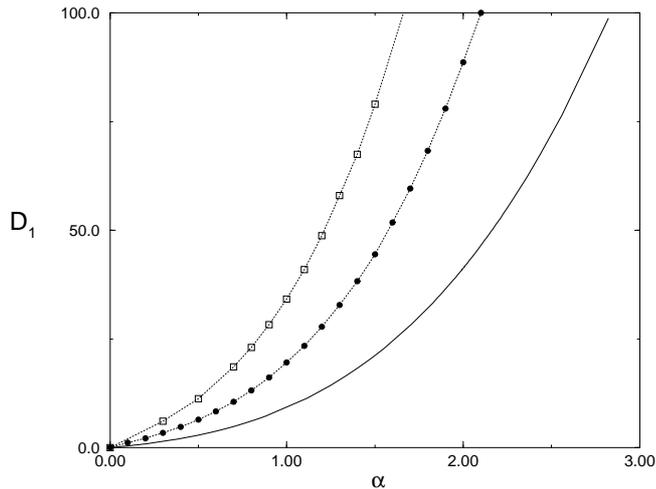}}
\caption{ Critical curves  obtained
 by   numerical simulations of  the single-well
  random  oscillator subject to multiplicative   Ornstein-Uhlenbeck noise. 
 The curves with black dots  ($\bullet$)
 and with squares ($\Box$)  correspond to  $\tau =1$ and  
   $\tau =2$, respectively. The full black line is  the  analytical result
 for the  white noise case.}
  \label{figpeyn}
\end{figure}

 From a physical point of view,
 the influence of a finite correlation time  on the shape of the
 critical curve  is an  interesting  open  question:
 due to the  finite  correlation time of the noise, the random oscillator 
  is  a  non-Markovian  
 random process and  there exists no  closed Fokker-Planck equation that 
 describes the dynamics of the Probability Distribution Function 
 (P.D.F.) in  the phase space. 
 This  non-Markovian feature hinders an exact solution in contrast with  the
 white noise case  where 
 a closed  formula for the  Lyapunov exponent was found.
  For the single-well stochastic  oscillator,
   we    have calculated \cite{kirPeyneau} 
 this Lyapunov exponent by using  different  approximations  and  have
 derived   the phase diagram.
  The main role  of the 
 correlation time of the noise, as can be seen in Figure \ref{figpeyn},
 is to enhance the stability region. When $\tau$ grows, amplitude
 of the noise required to destabilize the origin becomes bigger.
 This effect can be seen quantitavely in  the following exact 
 asymptotic expansions (see also \cite{crauel})~:
\begin{eqnarray}
 \hbox{ For small values of the noise amplitude~:}  \,\,\, 
  \alpha_c &\simeq&  \frac{\Delta  }{4 (1 + 4 \omega^2\tau^2)} \, , \\
 \hbox{ For large  values of the noise amplitude~:}  \,\,\,
   \alpha_c &\simeq& C  \Big( \frac{\Delta}{\omega\tau} \Big)^{1/4} \, , 
\end{eqnarray}
 where the constant $C$ is of order 1. Comparing with the white noise result
 we observe that the  behaviour of the critical curve 
  for large noise is modified in presence of time correlations;
 the asymprtotic exponent is 1/3 for white noise and 1/4 for colored
 noise even if the correlation time $\tau$ is small. The presence
 of a non-vanishing correlation time thus modifies the  scaling  
  characteristics of the system.  This change of scaling at large
 noise is related to the differentiability properties of the random
 potential as shown in \cite{delyon}~: 
 the white noise is continuous but nowhere
 differentiable whereas the  Ornstein-Uhlenbeck process has a 
 first derivative but no second derivative. 

\begin{figure}[th]
  \includegraphics[height=6.5cm]{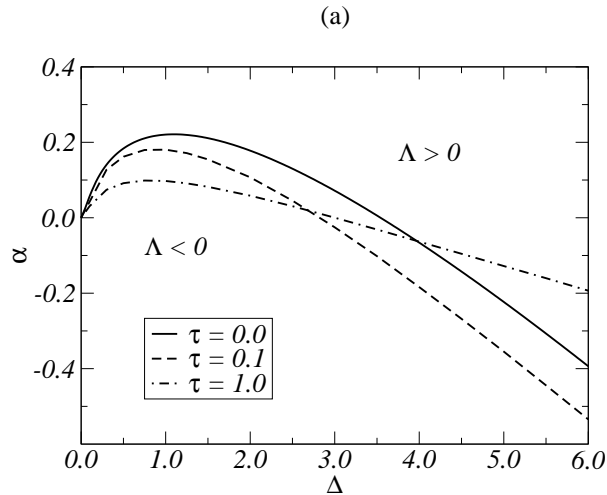}
 \bigskip
   \bigskip
   \caption{Phase diagram of the double-well oscillator subject
 to Ornstein-Uhlenbeck noise  with  correlation
time $\tau$. The  bifurcation lines for $\tau = 0.1$ and $1.0$ are obtained
from numerical  calculations  of the Lyapunov exponent. 
For comparison, we draw the (analytic)
white noise line. In this figure the dimensionless
 parameters  are given by  $ \alpha =  \frac{\omega^2}{\gamma^2} $
 and   $ \Delta   =  \frac{ {\mathcal D} }{ \gamma^3 }\,.$ }
  \label{fig:diagphase3}
\end{figure}

  For the double-well oscillator subject to  Ornstein-Uhlenbeck noise,
 the results of numerical computations of the phase diagram are
 shown in figure~\ref{fig:diagphase3}.
  Here also we observe that the noise-induced{\it stabilization} 
  of the deterministically unstable fixed point
 becomes less efficient in presence of time correlations. 
 In the weak noise limit, this  curve agrees with the prediction 
 of \cite{luecke1}:$\alpha_c(\Delta) \sim  \Delta/(2 (1 + \tau))$.
For a  noise amplitude of order 1, the bifurcation line
  is qualitatively similar 
to that obtained with white noise. However, depending on the value of
$\Delta$, the value of the bifurcation point $\alpha_c(\Delta)$
is not necessarily   a monotonic function of $\tau$.
 A precise  understanding  of this non-monotonous behaviour
 is lacking  and the 
 analytical theory of the stability of the
 double-well oscillator with time-correlated noise  still remains
 to be done.  Although  the obtention 
 of  exact results for the Ornstein-Uhlenbeck noise
 seems unlikely,  various recent  works \cite{vandenbroeck} indicate 
 that the Poisson process  provides a useful  model  
  for the study of time correlation
 effects in stochastic systems. The advantage of the Poisson process 
  is that it is amenable to exact analytic calculations.

 We conclude this section by emphasizing that
 the  results  obtained  here for the  oscillator with
 multiplicative noise could be used for other stochastic systems.
 For example, it has been shown by Schimansky-Geier et al. \cite{schimansky}
 that  the  random Duffing  oscillator with additive noise undergoes
 a  phase transition that does not manifest itself
 in the stationary P.D.F. (which is simply given by the
 Gibbs-Boltzmann formula). This  subtle  phase transition, which 
 affects the properties of the random attractor  of  motion
 in phase space,  can be formulated mathematically as a
 bifurcation  in an associated linear oscillator subject
 to a multiplicative  noise with a 
 a finite  correlation time.  If we approximate this noise  by an effective
  Ornstein-Uhlenbeck process, the system becomes identical to the
 one studied here.  

 \section{Scaling  behaviour  near the bifurcation  threshold} 
  \label{sec:scaling}

\begin{figure}[th]
 
  \includegraphics[height=6.5cm]{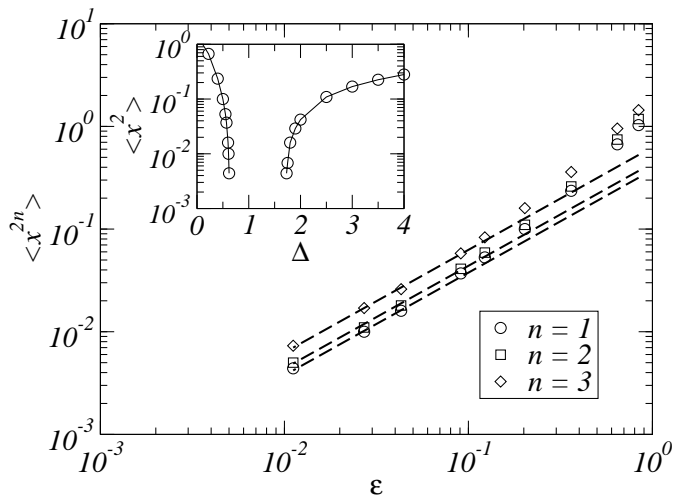}
\bigskip
\bigskip

 \caption{Behaviour of the moments near the
 reentrant  bifurcation threshold. The noisy  double-well  oscillator
subject to parametric white noise is simulated 
 with a fixed parameter $\alpha = 0.2$. The critical  value
  of the noise amplitude corresponding to the reentrant transition
 is found to be $\Delta_c \simeq 0.627$~: for  a noise amplitude
  $\Delta$ belonging to the interval (0.627--1.70) the origin
 is stabilized by the noise. 
  Even-order moments $\langle x^2 \rangle$, $\langle x^4 \rangle$
and $\langle x^6 \rangle$ are plotted versus the distance to threshold
$\epsilon = (\Delta_c - \Delta)/\Delta_c$ (symbols). Dashed lines 
respecting a linear behaviour $\langle x^{2n} \rangle \propto \epsilon$ are 
drawn to guide the eye. 
Inset: the mean square position (measured in the stationary regime) 
is non-zero for $\Delta < 0.627$ and  $\Delta > 1.70$, i.e., the stationary
 state is extended.} 
  \label{fig:mult}
\end{figure}

  For  deterministic Hopf bifurcation, the amplitude of 
 the order parameter $x$  exhibits a normal scaling
 behaviour in the vicinity of   the transition line; if $\epsilon$
 denotes the distance from threshold of the control parameter,
 we have in the deterministic case~:
\begin{eqnarray}
  \langle x^{2n} \rangle \sim \epsilon^n \, .
\label{Hopfnormal}
\end{eqnarray} 
  We now consider the case of  the stochastic oscillator.
 In figure~\ref{fig:mult}, we plot  the behaviour of the amplitude moments
 in the vicinity of the reentrant transition in a  
  double-well oscillator. The 
   damping rate $\alpha$  is fixed at  a   given value 
 and   the noise strength  $\Delta$   is chosen such
 that  $0 < \Delta -\Delta_c(\alpha) \ll 1$, where
 $\Delta_c(\alpha)$ is the critical value.  In this case, 
   the parameters of the  system 
 are tuned just   above the bifurcation threshold  and 
    the Lyapunov exponent  is slightly positive:  
 $0 < \Lambda  \ll 1$. We observe that 
 even-order moments
of the amplitude  scale linearly 
with the distance to threshold in the vicinity of the
 bifurcation line (note that odd-order moments  are 
equal to zero by symmetry)~:
\begin{eqnarray}
 \boxed{  \,\,\,\,   \,\,\,\,  \langle x^{2n} \rangle \sim \epsilon \, .
 \,\,\,\,   \,\,\,\, }
\label{Hopfstoch}
\end{eqnarray} 
 A similar scaling  behaviour is obtained by simulating the
 single-well oscillator or by replacing white noise by an
 Ornstein-Uhlenbeck process. Such a  behaviour 
   was also observed 
  in  random maps \cite{pikovsky} and  in  stochastic fields \cite{munoz2}.
 Such a strong multiscaling  seems to be a typical feature
 of  stochastic bifurcations and has been  noticed a long ago
 \cite{graham} in first
 order stochastic differential  equations. 
 The fact that the stochastic oscillator
  exhibits  a similar behaviour can be understood `physically'
 by noticing that the second order time derivative becomes irrelevant
 in the long-time limit.
   The mathematical explanation 
  for this behaviour is also  elementary \cite{philkir1}~: 
 near the bifurcation
 threshold, when the Lyapunov exponent $\Lambda \to 0$,
 the stationary  Probability Distribution Function $P_{{\rm stat}}$ exhibits a
 power law divergence at  the origin $(\dot x = x = 0)$,   and the region near
 the origin  dominates the statistics~: it can be shown
 that  near $E=0$,  the  stationary energy distribution scales as  
 $  P_{{\rm stat}}(E) \sim E^{c\Lambda -1} $ where $c$ is a positive
 constant. From this formula, one readily deduces that all
 moments of the energy (and therefore all moments of the amplitude
 of the oscillator)  are of the order $\Lambda$ which  itself grows  linearly 
 with $\epsilon$,  the distance from threshold.

 Another  remarkable feature of the order parameter near threshold
 is that the time series $ t \mapsto x(t)$ exhibits
 on-off intermittency. Again this  intermittent behaviour,
 first discovered in coupled dynamical systems \cite{yamada}
 and  in a system of
reaction-diffusion equations \cite{pikovskyonoff}, 
 is believed to be generic (see e.g.,  \cite{spiegel})
when an unstable system  is coupled to a system
that evolves in an unpredictable manner (multiplicative noise).

 Multiscaling  and intermittency  
 lead however to the following very   puzzling problems~:

\hfill\break
 {\bf Problem 1~:}
 The multiscaling  behaviour given in   equation~(\ref{Hopfstoch})
 is very clearly seen in computer experiments but does not
 seem to be observed in `real' experiments. Similarly,
It is surprising  that, despite the genericity of the
on-off intermittency mechanism (which  is  well established mathematically)
  this effect has scarcely  been
 reported  in experiments.  One might
  expect that any careful experimental
investigation of an instability should reveal on-off intermittency when
the system is close to the onset of instability, and is hence sensitive
to  unavoidable experimental noise in the control parameters.

\hfill\break
   {\bf Problem 2~:}  In well known works \cite{luecke1,luecke2}, 
 it   was predicted,   using perturbation theory, 
 that  the scaling of the stochastic bifurcation  should 
 be {\it the same} as that of the deterministic Hopf
 bifurcation, equation~(\ref{Hopfnormal}).
 This problem is analyzed in section~\ref{sec:Poincare}.

\hfill\break

\hfill\break

  Problem 1 is solved in  \cite{seb1,seb2}. 
  In these works,  it is shown that although the  noise  strength
 controls the transition between the absorbing and the oscillatory states, 
 the amplitude of the noise {\it is not  the only relevant}
 parameter responsible for multiscaling  and intermittency. 
 Rather, these effects are due to the  zero-frequency mode  of the noise;
 in other words  these effects depend on the  Power
   Spectrum Density (PSD) of the noise  and not  on its overall  amplitude.
   In order to identify which part  of the PSD
  of the random forcing really affects the
dynamics, one needs a random perturbation 
 with  a spectral density more complex than 
  that  of  white  noise or  of Ornstein-Uhlenbeck
 noise.  A useful type of noise   \cite{schimanskyharmonic}
 is the  {\it harmonic} noise $\zeta(t)$   whose
   autocorrelation function  is given by 
 \begin{eqnarray}
 \langle \zeta(t)\,\zeta(t+\tau) \rangle_s =  
  A  \exp(-2\pi \eta |\tau|)  \left(\cos(2\pi \Omega \tau) +
\frac{\eta}{\Omega} \sin(2\pi \Omega |\tau|)\right)  \, , \,\,\,
\,\,\, \label{selfcorrbr}
\end{eqnarray}
where $A$ is the noise amplitude;  the  corresponding  PSD  is  
 \begin{equation}
C(\nu)=\frac{A \eta (\Omega^2+\eta^2)}
{\pi[(\Omega^2+\eta^2-\nu^2)^2+2\nu^2\eta^2]}\,.
\label{spectrecol}
\end{equation}
 The value of the PDS at zero frequency is  therefore given by 
  \begin{equation}
S = A  \eta/\left[\pi(\eta^2+\Omega^2)\right]\, . 
\end{equation}
 For harmonic noise, the amplitude  $A$  and the
  zero frequency amplitude $S$ can be tuned {\it independently}. 

\begin{figure}[th]
  \includegraphics[height=7.0cm]{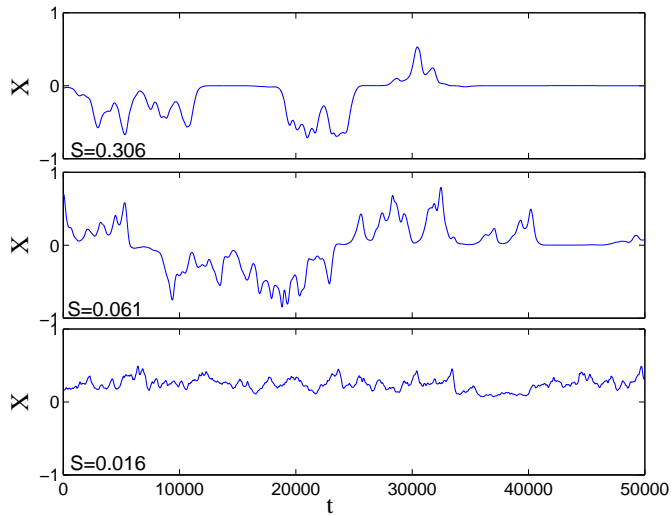}
   \caption{Temporal traces of the amplitude of a  Duffing oscillator subject
 to harmonic noise of constant amplitude $A=0.05$. 
  As  the  value $S$ of the zero frequency amplitude
 of the noise   is decreased from top to
bottom,  intermittent  behaviour is suppressed.}
  \label{fig:Intermittence}
\end{figure}

\begin{figure}[th]
   \includegraphics[height=7.0cm]{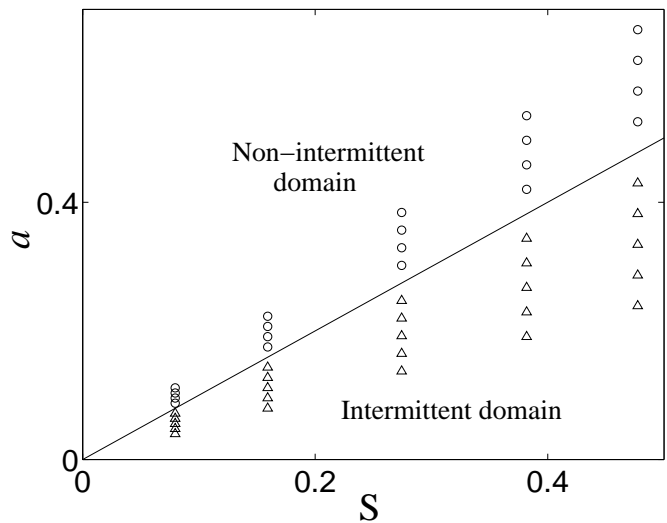}
    \caption{Intermittency phase diagram for  the 
 solution of  (\ref{eqbase}) with harmonic noise.
  ($\triangle$) : intermittent behavior, ($\circ$) : 
 non-intermittent behavior. The full line is the transition curve
  predicted by (\ref{eqcriterium}).}
   \label{fig:Dgphaseb}
 \end{figure}

 In figure~\ref{fig:Intermittence}, we plot 
  the time series of a   Duffing oscillator
   with a random frequency perturbation
 given by a  {\it harmonic} noise $\zeta(t)$. We observe
 that the  zero frequency amplitude $S$  is  the pertinent parameter
controlling the intermittent regime~:
   on-off intermittency disappears  when the value of $S$ is lowered.
 For the simple first-order model,
 \begin{equation}
 \dot{x}=(a+\zeta(t)) x-x^3\,,
\label{eqbase}
\end{equation}
 it is  possible to determine analytically 
 the transition line between intermittent
 behaviour and non-intermittent behaviour in the $(S,a)$ plane, using
 the cumulant expansion introduced by Van Kampen \cite{vankampen}.
 Intermittency  occurs when 
\begin{equation}
0<\frac{a}{S}<1     \,.
\label{eqcriterium}
\end{equation}
 This criterion is checked numerically in figure~\ref{fig:Dgphaseb}.

  Finally, it  is  possible to prove  that multiscaling  is a consequence
 of the intermittent behaviour of the order parameter $x(t)$ \cite{seb2}.
 Therefore as   the zero frequency noise amplitude  is reduced, 
 both   on-off intermittency and multiscaling are suppressed
 and normal scaling ~(\ref{Hopfnormal})  is recovered. 

   This  analysis  solves Problem 1 by 
  explaining  why many experimental investigations on the effect of a
  multiplicative noise on an instability do not display  on-off intermittency. 
 If the noise is high-pass filtered, as often required for  experimental 
 reasons, then the regime of intermittent behavior disappears.
  This is the case for instance in \cite{Francois2}: a ferrofluidic
  layer  undergoes the Rosensweig instability and peaks appear at the surface.
  The layer is then subject to a multiplicative noise through random
  vertical shaking. Close to the deterministic onset, the unstable
  mode submitted to a colored noise does not display intermittency.
  Another experimental obstacle to observe intermittency 
 is the presence of additive noise that destroys the symmetry of the system.

 \section{Remarks on the perturbative analysis of a noisy Hopf bifurcation}
\label{sec:Poincare}

 Problem 2 questions the relevance of perturbative expansions
 for studying noise induced bifurcations.  
   Consider, for example,  the  
   parametrically driven damped anharmonic oscillator  that  
 naturally appears in the study of many instabilities \cite{fauve}.  
 Such a system   is described  by the  following equation~:
\begin{equation}
   m  \ddot{x} + m \gamma \dot{x} =
  \left(\epsilon + \Delta \xi(t) \right) x -  x^3 \, ,  
 \label{eq:Lucke1}
\end{equation}
  where  $\epsilon$ is  the control parameter  and  the modulation
$\xi(t)$  is of arbitrary dynamics and statistics: it can be a
periodic function or  a  random noise.
   For small driving amplitudes
 $\Delta$,  L\"ucke and Schank   \cite{luecke1} have 
  performed   a Poincar\'e-Lindstedt expansion
 called the Poincar\'e-Lindstedt expansion 
 \cite{drazin,kevorkian,kleinert} (see \cite{khrustalev} for applications
 to field-theory). 
  They have  obtained  an expression 
 for the threshold   $\epsilon_c(\Delta)$  (at first order in
 $\Delta$). Their result 
 has been verified  both  numerically and 
 experimentally and is  also in agreement with
 the exact  result obtained for the  Gaussian white noise
 (in this case a closed formula is available for
 $\epsilon_c(\Delta)$   for  arbitrary values
 of $\Delta$).  Another result  obtained in   \cite{luecke1,luecke2}
 is the scaling of the moments near the threshold, 
\begin{equation}
  \langle x^{2n} \rangle = s_n 
\left( \epsilon - \epsilon_c(\Delta) \right)^{n} 
 + {\mathcal O}\left( ( \epsilon - \epsilon_c)^{n + 1}  \right)    \, ,
 \label{eq:Luckemoment}
\end{equation}
 where the constant $s_n$ depends on  $\xi(t)$ and on $\Delta$. The
 moments have   a  {\it normal scaling}  behaviour~:
 $ \langle x^{2n} \rangle$ scales as  $\langle x^2 \rangle^n$.   
 The bifurcation scaling exponent is equal to  1/2  and is
   the same as that of  a   deterministic Hopf bifurcation.  However,  
  this expression does not agree with the 
    results for   random iterated maps,    for 
  the random parametric  oscillator  and  
   with  recent  studies  on  On-Off intermittency 
   \cite{pikovsky, philkir1,seb1}.  These works 
  predict that   the variable   $x$  is   intermittent  and  that 
  the   moments of    $x$  exhibit  {\it  anomalous scaling}, 
\begin{equation}
   \langle x^{2n} \rangle  \simeq \kappa_n  (\epsilon - \epsilon_c) 
      \,\,\,  \hbox{  for all } \,\,\, n >0 \, ,
 \label{eq:anomalous}
\end{equation}
 {\it i.e.},  all the moments  grow linearly 
 with the distance from  threshold. This multiscaling  behaviour,
 confirmed by numerical simulations for a  Gaussian white noise,  
  was  derived using  effective  Fokker-Planck equations. 
 The origin of the  contradiction  between
  equations~(\ref{eq:Luckemoment})
  and (\ref{eq:anomalous})  lies in  the  divergences
 that appear in the Poincar\'e-Linsdtedt expansion. This fact,  
  identified  in  \cite{luecke2},  implies that   the results of
  \cite{luecke1} are valid only for  noises 
  that do  not have  low frequencies.

\subsection{A  simple model}

  The importance of    low frequencies in the noise spectrum
 can be seen analytically in a 
 a model  technically simpler 
   than  equation~(\ref{eq:Lucke1}).   
  To simplify our discussion we work on the first order  stochastic equation~:
\begin{equation}
    \dot{x} =
  \left( \epsilon   + \Delta \xi(t) \right) x -  x^{3} \, , 
\label{eq:Hopf}
\end{equation}
 that we already encountered  in the previous section. 
  Here,  the noise $\xi(t)$ is
  a Gaussian  stationary random process  with  zero mean value 
  and with a correlation function given by
\begin{equation}
  \langle  \xi(t)  \xi(t')  \rangle  = {\mathcal D}(|t - t'|) \, .
\label{eq:corr}
\end{equation}
The power spectrum of the noise, which is non-negative
 thanks to the  Wiener-Khinchin theorem \cite{vankampen}, is the Fourier
 transform of the correlation function
\begin{equation}
  \hat{\mathcal D}(\omega) =  \int_{-\infty}^{+\infty} {\rm d}t \exp(i\omega t)
    \langle  \xi(t)  \xi(0)  \rangle  = 
  \int_{-\infty}^{+\infty} {\rm d}t \exp(i\omega t)
{\mathcal D}(|t|) \, .
\label{eq:PSD}
\end{equation}

 Applying  elementary  dimensional analysis to   equation~(\ref{eq:Hopf}),
 we obtain  the  following scaling relations:
 \begin{equation}
  x \sim t^{{1}/{(2p)}} \, , \,\,\,\,\,\,\,\,
 \xi \sim t^{-1/2}  \, ,  \,\,\,\,\,\,\,\,
  \epsilon \sim \Delta^2 \sim t^{-1 } \, .
\label{eq:dimensions}
\end{equation}
The dimension of the noise  $\xi$ is so  chosen as to render the   
 power spectrum  $\hat{\mathcal D}(\omega)$  a dimensionless function.

 We first remark that the presence of noise does not modify
 the bifurcation threshold; indeed 
  the Lyapunov exponent of the system is given by  $\Lambda =  \epsilon $  
 and therefore the bifurcation always occurs at
  $\epsilon_c(\Delta) = 0$.

 When $\xi(t)$ is a Gaussian white noise, the  stationary
 solution  of the   Fokker-Planck equation
  corresponding to equation~(\ref{eq:Hopf}) is given by
  \begin{equation}
    P_{{\rm stat}}(x)  =  \frac{ 2} 
  {\Gamma( \alpha)  (\Delta^2)^{\alpha} } 
  x^{2\alpha  -1} 
  \exp\left(-\frac{x^{2}}{\Delta^2} \right)   \,\,\,  \,\,\,
{\rm  with } \,\,\,   \alpha =    \frac{\epsilon}{\Delta^2}   \, ,  
\label{eq:PDFblanc}
\end{equation}
   where $\Gamma$ represents the Euler Gamma-function.
 The bifurcation  threshold  is given by  $\epsilon = 0$; 
  for $\epsilon < 0 $, the solution~(\ref{eq:PDFblanc}) is not
  normalizable~:   the stationary PDF is the  Dirac distribution $\delta(x)$
 localized at the absorbing   fixed point  $x =0$. 
 For  $\epsilon >  0 $, the solution~(\ref{eq:PDFblanc}) is
 normalizable and   the moments of $x$  are given by 
  $  \langle x^{2n} \rangle  =    \frac{ \Gamma( \alpha + n) } 
  {\Gamma( \alpha) }   (\Delta^2)^{n}  \, . $ 
 In the vicinity of the threshold,  $\epsilon$ is small and 
 we recover multiscaling~:
  the moments scale linearly with  $\epsilon$, {\it i.e.},
 \begin{equation}
   \langle x^{2n} \rangle  \simeq  \epsilon  \,  (\Delta^2)^{n -1} 
  \Gamma(n) \, .
\label{eq:scalmomentblanc}
\end{equation}

\subsection{Exact solution for arbitrary noise}

  For an arbitrary noise $\xi(t)$, it is still  possible 
 to obtain an exact solution of  equation~(\ref{eq:Hopf})
 for all times. First, we shall
 derive some remarkable  identities  satisfied by the  exact solution. 

 We  define the differential operator ${\mathcal L}$ as follows,   
\begin{equation}
   {\mathcal L}  = \frac{ {\rm d}}{ {\rm d}t} - \Delta \xi(t) \, . 
\label{eq:defL}
\end{equation}
  Let us call $y_1(t)$ the solution
 of the adjoint equation  
 ${\mathcal L}^ \dagger  y_1  = 0 $, which is  given by
\begin{equation}
  y_1(t)  =    \exp\left( -  \Delta  \int_0^t \xi(u){\rm d}u\right) \, .
\label{eq:defy1}
\end{equation}
 Multiplying   equation~(\ref{eq:Hopf}) by the function 
   $y_1$  and taking average values,  we obtain 
\begin{equation}
 \langle  y_1 {\mathcal L} x   \rangle  =
 \langle     \epsilon  y_1  x  -  y_1  x^3 \rangle    \, .
\end{equation}
 Integrating  the left hand side of this equation by parts
 and taking  into account the fact 
  that $y_1$ is in the kernel of the adjoint 
  operator  ${\mathcal L}^ \dagger$, we derive  the following
  relation 
\begin{equation}
   \epsilon  = \frac{  \langle  y_1  x^3 \rangle }  
 {  \langle  y_1  x \rangle }  +
  \frac{ {\rm d}}{ {\rm d}t} \ln  \langle  y_1  x \rangle \, .
\label{eq:solvability}
\end{equation}
 Similarly, we have 
\begin{equation}
   \epsilon  =    \langle   x^2 \rangle   +
  \frac{ {\rm d}}{ {\rm d}t}   \langle  \ln (y_1  x) \rangle \, .
\label{eq:solvability2}
\end{equation}

   Dividing both sides of equation~(\ref{eq:Hopf}) by $x^3$ and
 using  the auxiliary variable $U = 1/x^2$, we observe that
  equation~(\ref{eq:Hopf}) becomes a {\it linear}  first order
 stochastic in $U$~:
\begin{equation}
  \frac{1}{2} \dot U =  1 - ( \epsilon + \Delta \xi) U \, .
\label{eq:varU}
\end{equation}
  This equation  can be solved exactly for all time
  by using the  method of variation
 of constants.
  Introducing
 the initial value $x(0) = \lambda$   that  has the dimensions 
\begin{equation}
    \lambda  \sim t^{{1}/{2}}  \, , 
\label{eq:dimlambda}
\end{equation}
we obtain the following explicit formula for $x$~:
\begin{equation}
      x(t) =  \frac{  \lambda  \exp\left( \epsilon t + \Delta   B_t \right)}
  { \sqrt{ 1 + 2 \lambda^2  \int_0^t  
\exp\left(2 \epsilon u + 2\Delta B_u \right) \rm{d}u  \, } \,\,\,\, } \, . 
\label{eq:formulex(t)}
\end{equation}
 We have defined here  the  auxiliary  random variable
 $B_t$~:
 \begin{equation}
       B_t  =   \int_0^t \xi(u){\rm d}u \, . 
  \label{eq:defB}
\end{equation}
   Because  $\xi$ is  taken to be a  Gaussian random process,
  $B_t$ is also Gaussian.

\subsection{Importance of  low frequencies in  the noise spectrum}

   In order to determine  the scaling on the moments of $x$,  we must
 evaluate expressions of the type 
 $\langle \exp \Delta B_t  \rangle$.
 Because $B_t$ is a Gaussian random variable, this quantity is given
 in terms of  the variance of $B_t$~:
\begin{eqnarray}
     \langle B_t^2 \rangle  =
   \int_0^t\int_0^t  \langle \xi(u) \xi(v) \rangle{\rm d}u {\rm d}v 
   =    \int_0^t\int_0^t   {\mathcal D}(|u-v|) \rangle{\rm d}u {\rm d}v 
   =   \int_{-\infty}^{+\infty} 
 \frac{ 1 - \cos\omega t}{\omega^2} \,\,
   \frac{ \hat{\mathcal D}(\omega) {\rm d}\omega}{\pi}\, .
  \label{eq:varB}
\end{eqnarray}
The last integral is well defined at $\omega  = 0$ (the time $t$
 introduces an effective low frequency  cut-off for $\omega \sim 1/t$).
 The behaviour of  $\langle B_t^2 \rangle$ for  $t \to \infty$
 depends on the  behaviour of  $\hat{\mathcal D}(\omega)$ at 
 $\omega  \to  0$. The following  two cases must be distinguished~:
 
 (i) The spectrum of the noise vanishes at 
  low frequencies, {\it i.e.},
 ${\mathcal D}(0) = 0$. Because  $\hat{\mathcal D}(\omega)$
 is an even function of $\omega$,  we suppose that
  $\hat{\mathcal D}(\omega)  \sim \omega^2 $ for
  $\omega \sim 0$ (we disgard  non-analytic
 behaviour of the power spectrum at the origin. Such   non-analyticity  would
 correspond to long tails in the correlation function of the noise). 

 (ii) The power spectrum of the noise is finite at  $\omega =  0$,
 {\it i.e.},  ${\mathcal D}(0) > 0$.

 In case (i), the long time limit of equation~(\ref{eq:varB}) 
 is readily derived and we obtain (by using the Riemann-Lebesgue
 lemma)
\begin{equation}
    \langle B_t^2 \rangle   \rightarrow  \int_{-\infty}^{+\infty} 
   \frac{ \hat{\mathcal D}(\omega) {\rm d}\omega}{ \pi \, \omega^2} \,\,\,\,\,
   \hbox{ when } \,\,\,\,\, t \to \infty \,. 
\end{equation}
The variance of $B_t$ has  a {\it finite} limit at large 
 times. 

 In case (ii), the integral on the right hand side of equation~(\ref{eq:varB})
 diverges when  $ t \to \infty$ and its leading behaviour is
\begin{equation}
 \langle B_t^2 \rangle  =  \frac{ t }{\pi } \int_{-\infty}^{+\infty} 
 \frac{ 1 - \cos u }{ u^2} \,\,
    \hat{\mathcal D}\left(\frac{u}{t}\right) {\rm d}u
   \rightarrow    \hat{\mathcal D}(0) t    \, .
\end{equation}
The variance of $B_t$  grows linearly with time in the long time
 limit.

\subsection{Behaviour of the moments}

   We have seen that the behaviour of the variance of $B_t$ in the 
 long time limit crucially depends on the on the behaviour
 of the noise spectrum at low frequencies. When 
 ${\mathcal D}(0) > 0$  the noise $\xi(t)$ acts dominantly
  as a white noise and the multiscaling behaviour~(\ref{eq:scalmomentblanc})
 is recovered when $t \to \infty$.

 When ${\mathcal D}(0) = 0$, the random variable $B_t$
 has a finite variance even when  $t \to \infty$. 
  Using equation~(\ref{eq:formulex(t)}) and keeping
 only the dominant terms in the long time limit, we obtain 
\begin{equation}
      x(t) \simeq 
  \frac{    \exp\left( \epsilon t + \Delta   B_t \right)}
  { \sqrt{ 2   \int_0^t  
\exp\left(2 \epsilon u + 2\Delta B_u \right) \rm{d}u  \, } \,\,\,\, } \, . 
\label{eq:formule2x(t)}
\end{equation}
  Therefore, we have
 \begin{equation} 
    x^{2n}(t)    \simeq 
  \Big( \,\,\, 
 2 \int_0^t \exp\left(-2 \epsilon (t-u) + 2\Delta (B_u- B_t) \right)
   \rm{d}u   \,\,\,   \Big)^{-n}  
\label{eq:xpower2n} 
\end{equation}
 The fluctuations of the random variable being of order one, 
 the integral on the  r.h.s. of this equation is dominated by
 the contribution of the linear term $ -2\epsilon (t-u)$. The contribution
  is maximal in the region   $u \simeq t$. Therefore,  we have
\begin{equation}
   \langle x^{2n} \rangle \simeq  C_n \epsilon^n  \, , 
\label{eq:scalmom}
\end{equation}
 where the constant $C_n$ depends a priori on the statistical properties
   of the noise. 
   This equation predicts that subject to  a noise
  without  low  frequencies, the amplitude $x(t)$ exhibits   a   normal 
  scaling behaviour identical to that  of  the deterministic case.

 \section{Conclusion}

  The  stochastic  oscillator is an ideal model to study the effect
  of random perturbations on a  nonlinear dynamical
  system. Various effects of noise can be demonstrated on this
  simple model~: noise can shift bifurcation thresholds,
  can  create  new  phases   by destabilizing (resp. stabilizing)
  stable (resp. unstable)  fixed points, can induce
  reentrant behaviour. 
 The relevant parameter that determines the long time
 behaviour of the system is the Lyapunov exponent
 of the underlying linearized dynamics. 
 Exact results can be derived for
  white noise and for dichotomous Poisson noise.
 When
 the system is coupled to the smooth Ornstein-Uhlenbeck
 process, analytical treatments have to rely on various
 approximations, however qualitative aspects on time-correlations
 are well understood.

  The effect of noise on the  scaling behaviour is subtle~:
 intermittency and multiscaling  can appear if the relative
 weight of the zero-frequency mode of  noise is large enough.
 If low frequencies are filtered out, the noise-induced bifurcation
 becomes qualitatively similar to the deterministic transition.
 These features are difficult to extract from a  perturbative
 analysis of the Langevin equation~:
  they appear as divergences in the perturbative expansion
 which must be resummed in order to get correct results. Indeed, 
  multiscaling   of the order parameter can not be revealed
 from   a finite  order truncation of the  expansion. 

 Despite the formal simplicity of the problem, exact
 results have  been   derived  only  recently and many 
  questions still remain to be addressed.  For example,  little
 is known about the  relevance
 of the noise spectrum on barrier crossing problems and on stochastic
 resonance. 
  Analytical results  
 for  higher dimensional systems and for  stochastic fields
 are  rare~:  such results  could be relevant for 
 the study of the trapping of quasi one-dimensional
 Bose-Einstein condensates in random potentials \cite{aspect}.

\end{document}